\documentclass[usenatbib,usegraphicx,letterpaper]{mn2e}
\usepackage{amssymb}
\usepackage{epsfig}
\usepackage{amsmath}
\usepackage{color}
\usepackage{float}
\usepackage{natbib}
\usepackage{graphicx}

\begin{document}

\title[Probing Satellite Quenching]{Probing Satellite Quenching With Galaxy Clustering} 
%\author[Chamberlain et al.]{Robert T. Chamberlain\altaffilmark{1},  Neal Dalal\altaffilmark{1,2},  Andrew Hearin\altaffilmark{3}, and Paul Ricker\altaffilmark{2}} 
\author[Chamberlain et al.]{Robert T. Chamberlain$^{1}$,  Neal Dalal$^{1,2}$,  Andrew Hearin$^{3}$, and Paul Ricker$^{2}$\\
$^{1}${Department of Physics, University of Illinois, Urbana, IL 61801}\\
$^{2}${Department of Astronomy, University of Illinois, Urbana, IL 61801}\\
$^{3}${Yale Center for Astronomy \& Astrophysics, Yale University, New Haven, CT}}

\maketitle

%\altaffiltext{1}{Department of Physics, University of Illinois, Urbana, IL 61801}
%\altaffiltext{2}{Department of Astronomy, University of Illinois, Urbana, IL 61801}
%\altaffiltext{3}{Yale Center for Astronomy \& Astrophysics, Yale University, New Haven, CT}

\begin{abstract} Satellites within simulated massive clusters are significantly spatially correlated with each other, even when those satellites are not gravitationally bound to each other.  This correlation is produced by satellites that entered their hosts relatively recently, and is undetectable for satellites that have resided in their hosts for multiple dynamical timescales.  Therefore, a measurement of clustering statistics of cluster satellites may be used to determine the typical accretion redshifts of those satellites into their observed hosts.  We argue that such measurements may be used to determine the fraction of satellite galaxies that were quenched by their current hosts, thereby discriminating among models for quenching of star formation in satellite galaxies.
\end{abstract}

%\keywords{cosmology: dark matter - galaxies: halos - methods: statistical, N-body simulations}

\section{Introduction}

Astronomical surveys have long established a correlation between galaxy properties and environmental density \citep[e.g.,][]{oemler74,DavisGeller1976,Dressler1980,PostmanGeller1984}.   At fixed luminosity or stellar mass, high-density environments like groups and clusters exhibit a higher fraction of quiescent galaxies, whereas more isolated environments have a larger fraction of galaxies that are actively star-forming \citep[e.g.,][]{Balogh1997,Balogh2004,Kauffmann2004,Blanton2005}.  For example, among galaxies in the local universe with stellar mass near $10^{10} M_\odot$, the red fraction within massive groups and clusters is $\sim70\%,$ while in the field, the red fraction is far smaller, closer to $\sim25\%$ \citep{Weinmann2006}.  This relationship is known to persist at least as far back as $z\sim1$ \citep{Cucciati2006,Cooper2007,Peng10}, over seven billion years into the cosmic past. 

Historically, proposals for the physical origin of these phenomenological trends have focused on processes that are specific to the interior regions of particularly massive dark matter halos. For example, the hot, diffuse gas belonging to a satellite may become unbound as the galaxy orbits in the potential well of its host halo \citep{Larson1980,Balogh2000}, resulting in a gradual attenuation of star formation dubbed ``strangulation'' in \citet{BaloghMorris2000}. In a more rapid process proposed in \citet{1972GunnGott}, ram-pressure experienced by a satellite galaxy orbiting in the hot, virialized gas of its host halo has the potential to directly strip the satellite of its cold gas reservoir \citep[see also][for investigations of this phenomenon in hydrodynamical simulations]{Abadi1999,Quilis2000}. Satellite evolution may also be influenced by the cumulative effect of many high-speed encounters with other satellites, a process referred to as ``harassment'' \citep{FaroukiShapiro1981,Moore1996}. 

Early attempts at the semi-analytical modeling of galaxy evolution (SAMs) also focused on the extreme environmental conditions inside massive host halos. For example, the SAMs introduced in \citet{Kauffmann1993} and \citet{Cole1994} assumed instantaneous stripping of a satellite's gas upon infall, resulting in satellites that were predicted to be significantly redder than those observed in the local universe \citep{Weinmann2006,Baldry2006}.
Updates to these early models have consistently involved more gradual implementations of satellite-specific processes, and have brought predictions into broadly better agreement with observational data \citep[see, e.g.,][]{Font2008,Weinmann2010,Guo2011,Somerville2012}. However, significant tension between the predicted and observed satellite trends remains \citep{Lu2013,Kauffmann2013}, and there are several indications that this tension is due, at least in part, to an overestimate of the efficiency of intra-host processes \citep{Kimm2009,Wang2014,Hearin2014}.

Emphasis on the host halo playing the dominant role in satellite quenching can also be seen in contemporary empirical models of galaxy evolution. In the model introduced in \citet{Wetzel2013}, satellite galaxies evolve as centrals until the time $t_{\mathrm{inf}}$ that the satellite first passes within the virial radius of some larger host halo; after a delay time $t_{\mathrm{delay}}$ referred to as the ``quenching timescale'', the star formation rate in the satellite decays exponentially. Using a group catalog \citep{Tinker2011} constructed from a volume-limited sample of SDSS galaxies \citep{2000AJ....120.1579Y} with $M_{\ast}>10^{9.7}M_{\odot},$ the authors tuned the value of $t_{\mathrm{delay}}$ so that the model reproduced the satellite quenched fraction, and its variation with host-centric distance, as a function of the halo mass of groups, rich groups, and clusters. The quenching timescale $t_{\mathrm{delay}}$ must be a significant fraction of the Hubble time (6-9 Gyrs) to explain the distribution of low-mass quiescent satellites, a result that has been confirmed by multiple groups \citep[e.g.,][]{DeLucia2012,Wheeler2014}.

Although intra-host quenching is a natural explanation for the observed relation between local environment and quiescent fraction, quenching of star formation might not necessarily be driven by the halo environment.  Recently, \citet{Watson2014} have argued that it is possible to reproduce many statistical properties of galaxies, including the relation between local density and red fraction, using the `age-matching' prescription of \citet{Hearin2013}.  As shown in \citet{Hearin2013b}, age matching is a variant of subhalo abundance matching \citep{Conroy2006}, in which dark matter halos and subhalos are associated with galaxies with stellar populations whose ages are related to the assembly history of the (sub)halos.  At fixed stellar mass, quiescent galaxies are placed in relatively older halos than actively star-forming galaxies.  This correspondence between stellar age and halo age could arise if the transition from fast growth to slow growth in halos cuts off the supply of gas needed to form stars \citep{Feldmann2014}.  

The \citeauthor{Hearin2013} prescription quantitatively reproduces the correlation between quiescent fraction and local density, without invoking any physics related to the infall of satellites into larger halos. In this model, environmental trends arise because (sub)halos in high density environments tend to assemble their mass earlier than (sub)halos in low density environments. And yet, the age matching prescription is able to reproduce the same observational statistics deriving from galaxy group catalogs as those used to fine-tune the empirical and semi-analytic models discussed above. This directly implies that traditional group-based measurements are incapable of discriminating between  models predicated upon radically different assumptions, and highlights that the well-known correlation between halo assembly and environment \citep[e.g.,][]{GaoWhite2005,Dalal2008} represents a fundamental degeneracy in satellite quenching models. This degeneracy is also unbroken by the two-point correlation function split on broadband color, which limits the ability of this traditional statistic to robustly constrain models of galaxy evolution, particularly satellites \citep[see][figure 11]{Zentner2013}.

Fortunately, it is possible to distinguish observationally between these two classes of scenarios.  If infall into massive halos is responsible for all quenching of star formation in the galaxies inside massive groups and clusters, then most of the cluster galaxies that have been red and dead since high redshift (e.g. $z\sim1$) have been inside massive halos since $z\gtrsim1$.  While inside those massive halos, satellite galaxies experience significant tidal forces.  In this paper, we argue that tidal effects can provide a simple signature of the amount of time that a population of galaxies has typically resided inside of their host halos.  More specifically, measurements of satellite clustering can be used to determine the fraction of satellites that have been in their current hosts for multiple dynamical times.  If quenching is indeed a slow process that requires massive hosts, such measurements can place interesting constraints on quenching models.

\section{Galaxy clustering beyond the tidal radius}

\begin{figure*}
\centerline{\includegraphics[width=0.45\textwidth]{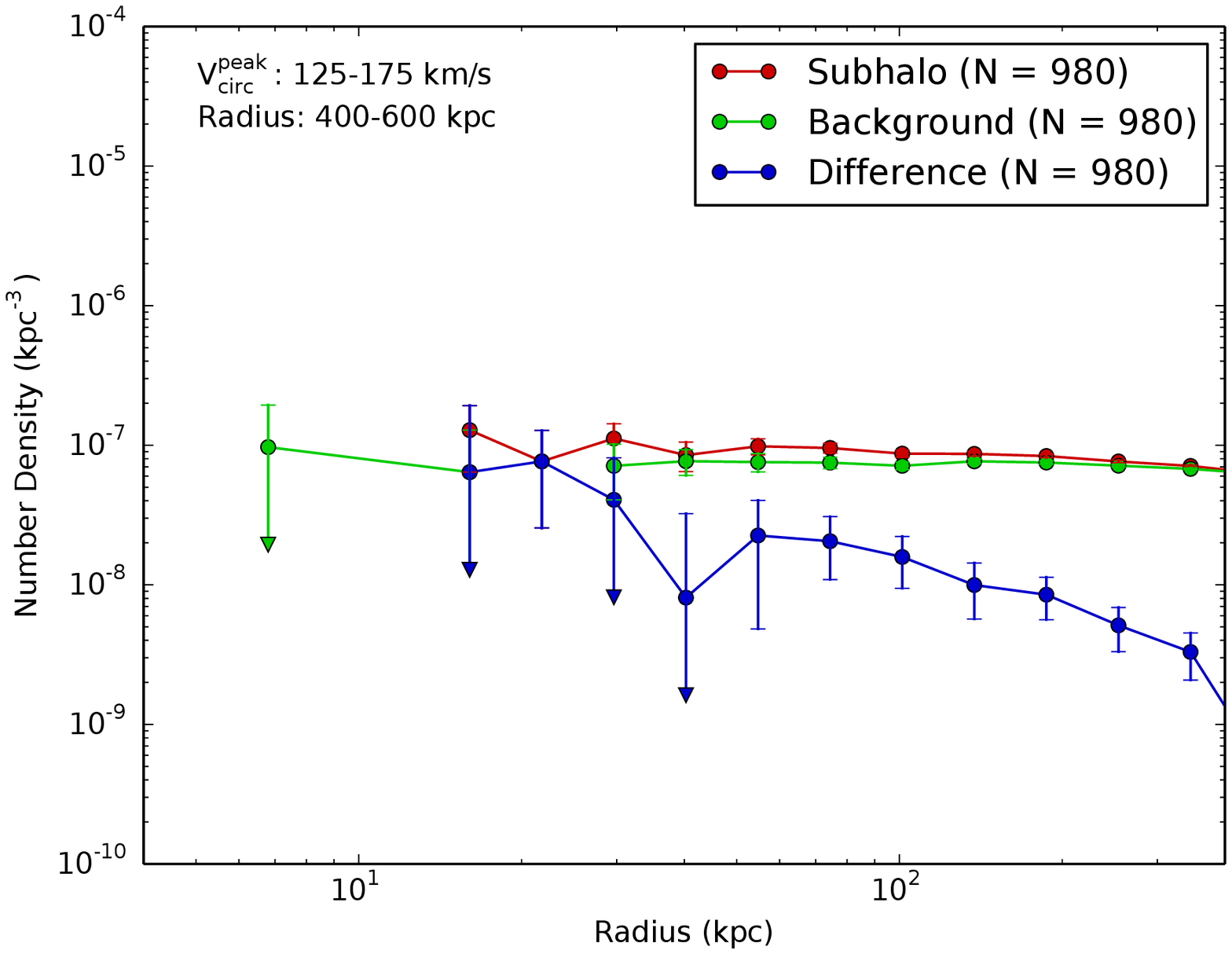}
\quad\includegraphics[width=0.45\textwidth]{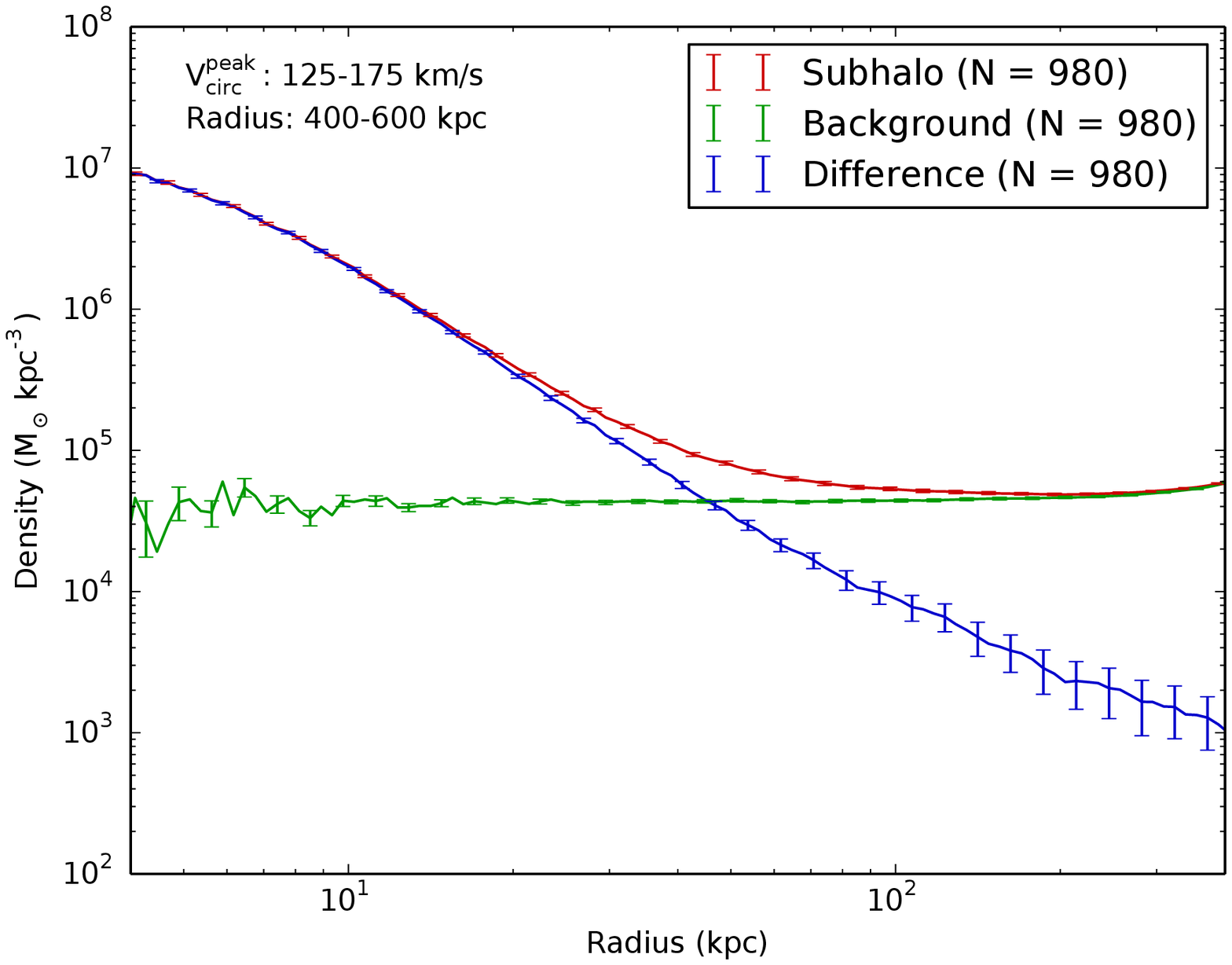}}
\caption{Stacked profiles centered on Rockstar subhalos with 125 km/s $< V_{\rm peak} <$ 175 km/s, in hosts of mass $10^{14} M_\odot < M_{\rm host} < 2\times 10^{14} M_\odot$, at cluster-centric radii of 400 kpc $< R <$ 600 kpc.  (Left) The average number density of other subhalos as a function of distance from the stacked subhalos.  The red curve shows the average profile centered on subhalos, and the green curve shows the background, estimated from the average profile centered on the antipodal point for each subhalo.  The blue curve shows the difference between red and green, corresponding to the subhalos that are spatially correlated with the stacked subhalos.  In the legend, $N$ indicates the number of subhalos that were stacked.  Downward pointing arrows indicate upper limits, i.e.\ points that are 1-$\sigma$ consistent with zero.  (Right) Similar to the left, but now plotting average mass density profiles instead of number of neighbors.  The tidal radius is expected to fall somewhere near the intersection of the blue and green curves, around 50 kpc in this case.  Therefore, material at even larger distances (e.g. $r>100$ kpc) is unbound to these subhalos.
\label{fig:allstack}}
\end{figure*}

Galaxies are spatially correlated with each other \citep[e.g.][]{Peebles}, a fact that is also true of the galaxies that fall into massive clusters and become satellite galaxies.  For example, a significant fraction ($\approx 30-40\%$) of galaxies enter clusters via bound groups \citep{Berrier2009,McGee2009}.  Once inside the clusters, tidal gravity can unbind these groups, and once galaxies become unbound from each other, their spatial correlations decay on a timescale of order a few dynamical times.

We might naively expect that satellite galaxies inside massive groups and clusters would exhibit little if any spatial correlations with each other past their tidal radii.  This would be incorrect, however, because a large fraction of cluster galaxies have not been inside their hosts for many dynamical times \citep{White2010,Cohn2012}.  This is because the dynamical time inside a halo can be a large fraction of the Hubble time (especially for large cluster-centric radii, $r\sim r_{\rm vir}$), and also because massive clusters tend to assemble much of their mass relatively late, $z < 1$.  Therefore, the spatial correlations of many satellite galaxies at separations $r>r_{\rm tidal}$ do not have sufficient time to decay completely, leading to observable spatial correlations of galaxies within clusters \citep{Cohn2012}.

To illustrate this, we plot in Figure \ref{fig:allstack} the 2-point correlation function of selected subhalos in the Bolshoi simulation \citep{2011ApJ...740..102K,2013AN....334..691R}.  Halos and subhalos were obtained from the publicly available Rockstar \citep{2013ApJ...762..109B} catalogs and merger trees available at the Multidark website\footnote{www.multidark.org}.  Following \citet{Reddick2013}, we stack subhalos of similar peak circular velocity, under the assumption that $V_{\rm peak}$ is the subhalo property that correlates most tightly with stellar mass.  We selected samples of subhalos in bins of $V_{\rm peak}$ and cluster-centric radius $R$, and then computed the average number density $\bar n(r)$ of neighboring subhalos as a function of separation $r$, as well as the average mass density profile $\bar\rho(r)$.  Uncertainties in the stacked profiles were estimated using bootstrap errors.  Note that because these are correlation functions in real space, the errors in the profile are themselves highly correlated, which is why the profiles may appear smoother than one might expect from the size of the error bars.  Also note that this figure plots the clustering of subhalos relative to each other; this is distinct from the radial profile of satellites within clusters.  The latter corresponds to the satellite-cluster 2-point correlation function, whereas Figure \ref{fig:allstack} plots the subhalo-subhalo correlation function within cluster-sized halos.  

The stacked profile plotted in Figure \ref{fig:allstack} requires some explanation, since it contains two distinct contributions.  First, there is a contribution associated with the mean radial profile of the host cluster.  When we stack on galaxies at cluster-centric radii $R$, then we would expect the average number of other nearby satellites at distances $r\lesssim R$ to be about $n(R)$, nearly independent of $r$.  We can think of this as the `background' density of galaxies at cluster-centric radius $R$.  In addition to this term, however, there are also possible satellite-satellite pairwise correlations.  For example, two sub-subhalos that are both bound to the same substructure will be spatially correlated with each other, above and beyond the correlations provided by the host profile $n(R)$.  It is this second contribution that we are interested in, since tidal gravity will tend to diminish such correlations for satellites that are not bound to each other.  We would therefore like to subtract the background contribution to the stacked profile, in order to measure the excess correlation of unbound material.  One simple way to estimate the background, suggested by \citet{PastorMira2011}, is to compute the average profile stacked on the antipodal point to each subhalo, i.e.\ the point at cluster-centric radius $R$ that is exactly opposite to the subhalo, relative to the host centroid.  The difference between the subhalo stacked profile and the opposite stacked profile reveals the excess correlations of subhalos with each other.

As Figure \ref{fig:allstack} illustrates, there are significant correlations of subhalos persisting to large radii.  Similar behavior holds true for the mass correlated with subhalos.  The average mass profiles were computed in a similar way as the average number profiles, by stacking dark matter particles as a function of distance relative to subhalos.  Because there are many more particles than subhalos, the statistical errors on the stacked mass profiles are far smaller than the errors on the satellite profile.  In both cases, there is a clear, significant and positive excess of mass persisting to large radius.  From the stacked mass profiles, we can also roughly estimate the approximate tidal radius: the Hill radius will occur where the density associated with subhalos (blue curve) becomes comparable to the background density (green curve).  For the subhalo sample plotted in Figure \ref{fig:allstack}, the Hill radius is of order 50 kpc, implying that the excess correlations seen at much larger radii are with material that is unbound.

\begin{figure*}
\centerline{\includegraphics[width=0.46\textwidth]{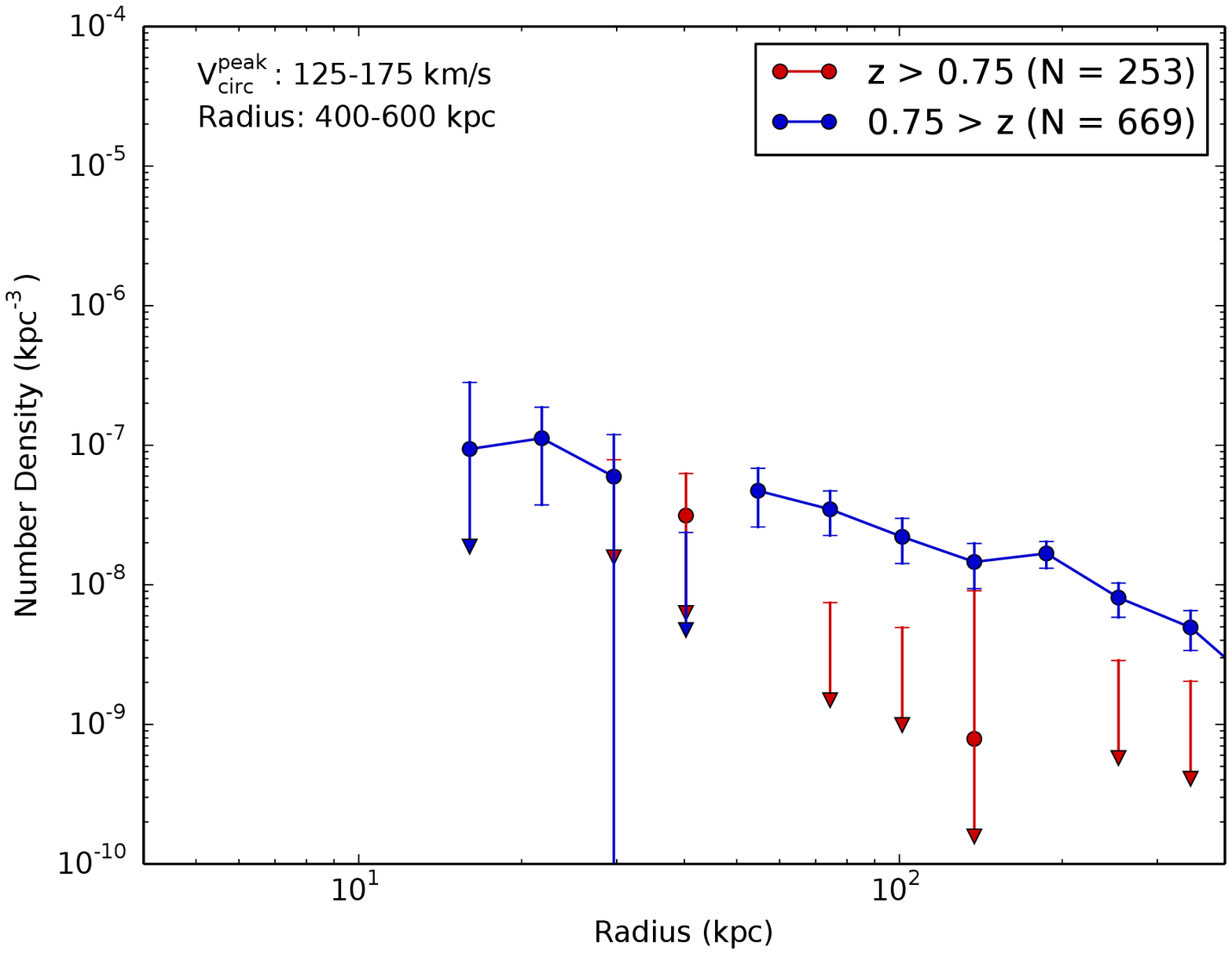}
\quad\includegraphics[width=0.45\textwidth]{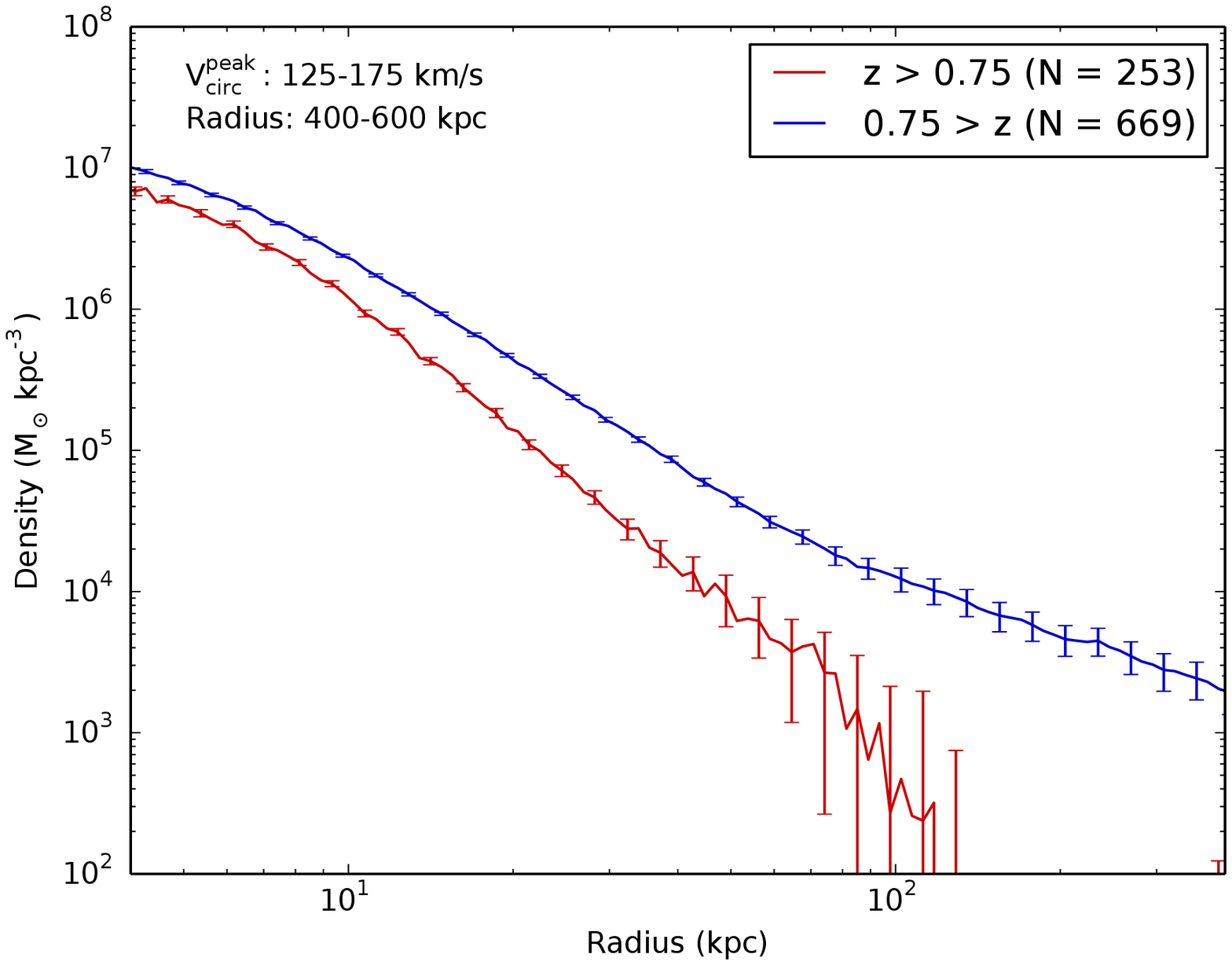}}
 \caption{Difference profiles for the same subhalos as in Figure \ref{fig:allstack}, split on accretion redshift.  The left panel shows the average number of neighboring subhalos, and the right panel shows the average mass density profile.  The oldest subset shows little correlated unbound mass, in contrast to the remainder of the subhalos.  The number density profiles appear consistent with this behavior, albeit with larger errors due to the smaller number of subhalos compared to dark matter particles in the simulation.  In the left panel, downward triangles indicate data points consistent with zero at 1-$\sigma$.  The oldest subhalos (red points) exhibit no detectable clustering at large radius, in contrast to the younger subhalo sample (blue points) at similar radii. 
\label{fig:acc}}
\end{figure*}

The presence of significant correlations with unbound material implies that many of the subhalos in our cluster sample have not been inside their host halos for long, compared to the dynamical time.  We can verify this by splitting our subhalo sample based on age.  From the publicly available merger trees for Rockstar subhalos provided by \citet{2013ApJ...763...18B}, we can determine the redshift at which each subhalo entered its $z=0$ host halo; we denote this redshift as $z_{\rm acc}$.  We then split the subhalo samples into groups of different $z_{\rm acc}$, and plot their respective stacked profiles in Figure \ref{fig:acc}.  As the figure illustrates, the oldest subsample of satellites, with $z_{\rm acc}>0.75$, indeed exhibit no significant correlations with mass at large distances well beyond the estimated Hill radius.  The excess observed in the combined sample (Figure \ref{fig:allstack}) is driven by the more recently accreted subhalos.  If we estimate the dynamical time as the orbital timescale for a circular orbit at radius $r$, then $t_{\rm dyn}=2\pi r/v_{\rm circ}(r) = 2\pi (GM/r^3)^{-1/2} \approx (\pi/G\rho)^{1/2}$, where the last approximate equality is only valid for nearly isothermal profiles, $\rho\propto r^{-2}$.  For the background density $\rho\approx 3\times 10^{13} M_\odot/{\rm Mpc}^3$ at this location ($R\sim 500$ kpc), the dynamical time is $t_{\rm dyn}\approx 5$ Gyr, corresponding to a lookback redshift $z\sim 0.5$.  As we can see, subahalos that have been inside their hosts for only $\lesssim 1$ dynamical times retain significant correlations with unbound material.

\section{Is quenching associated with infall?}

The result shown in Figure \ref{fig:acc} suggests an immediate test of environmental quenching models for satellite galaxies.  By measuring 2-point statistics of satellites within host halos, we can determine the fraction of those satellites that have been inside their hosts for multiple dynamical times.  By examining regions within clusters where the local dynamical time is less than the $\gtrsim 6$ Gyr quenching timescales found by previous analyses, we can thereby determine the fraction of quiescent satellites that were quenched by their current hosts.  This fraction is expected to be large, especially for satellites in massive groups and clusters with $M_{\rm host} \sim 10^{14} M_\odot$ \citep{Wetzel2013}.  If a large fraction of quiescent satellites have been inside their hosts for long amounts of time, then their 2-point correlations will resemble the oldest subset shown in Figure \ref{fig:acc}.  Conversely, if quenching of star formation is completely unrelated to infall into massive halos, then satellites with old or young stellar populations will not necessarily show significantly different correlations with unbound material.

\begin{figure*}
\centerline{\includegraphics[width=0.31\textwidth]{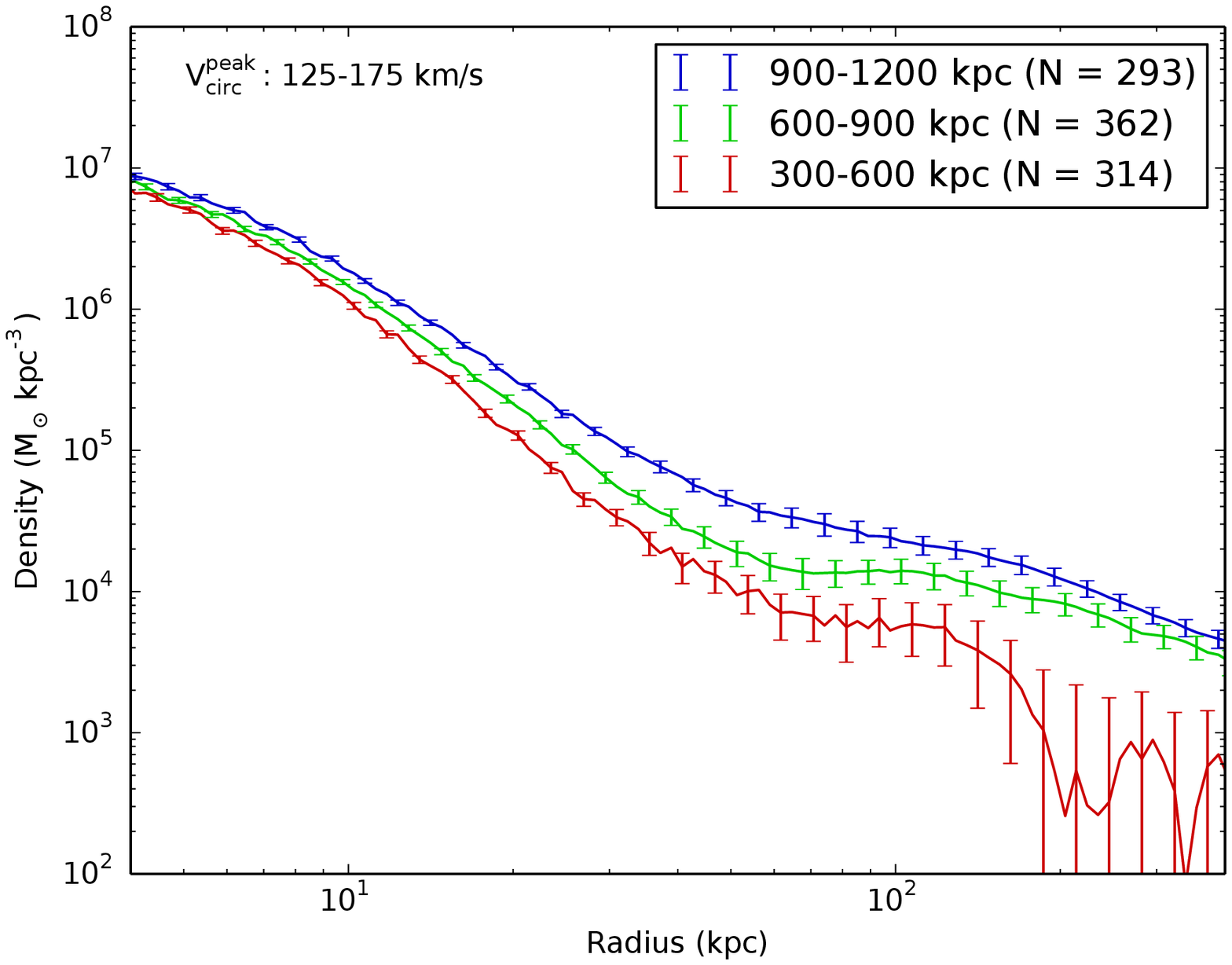}
\quad\includegraphics[width=0.31\textwidth]{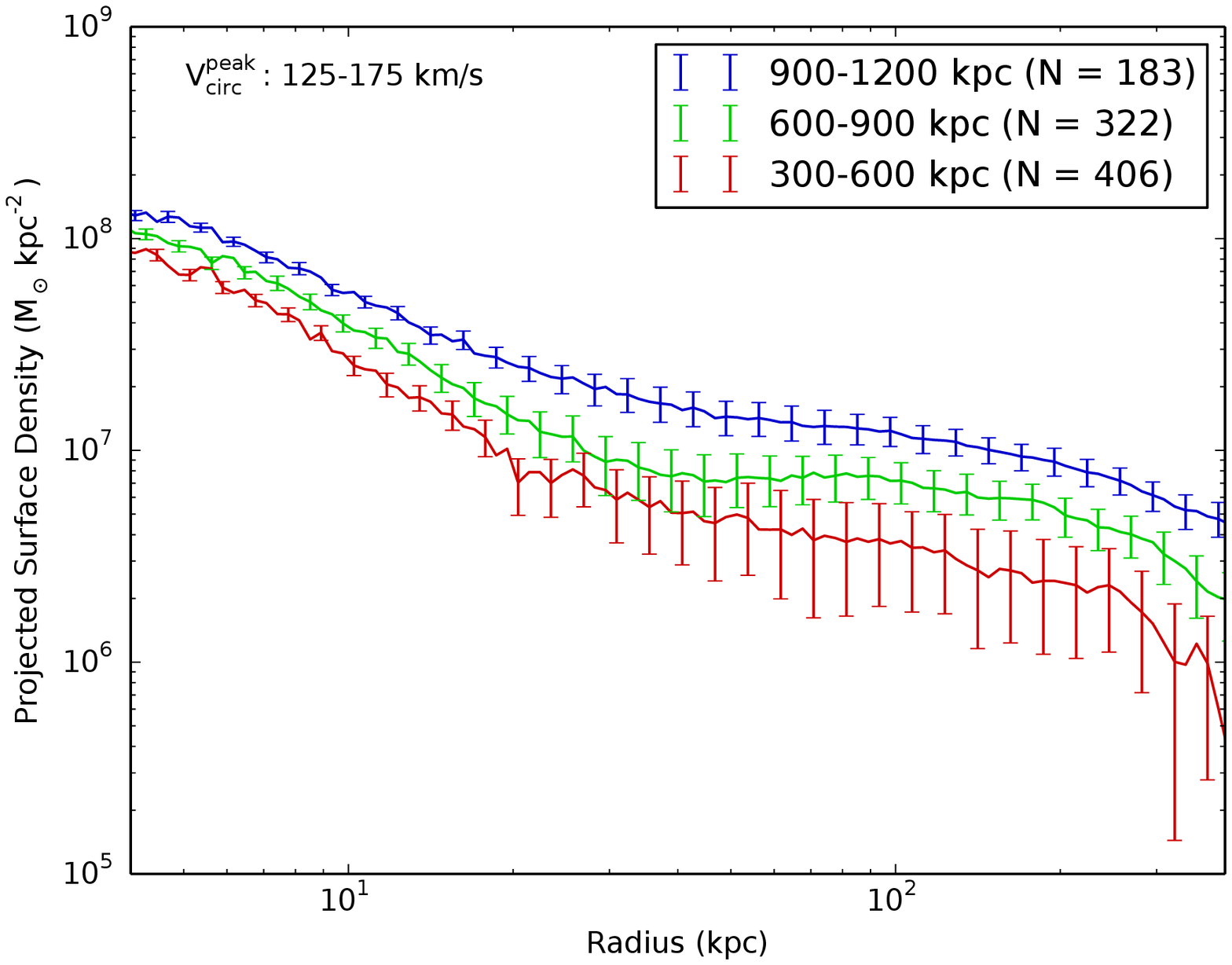}
\quad\includegraphics[width=0.31\textwidth]{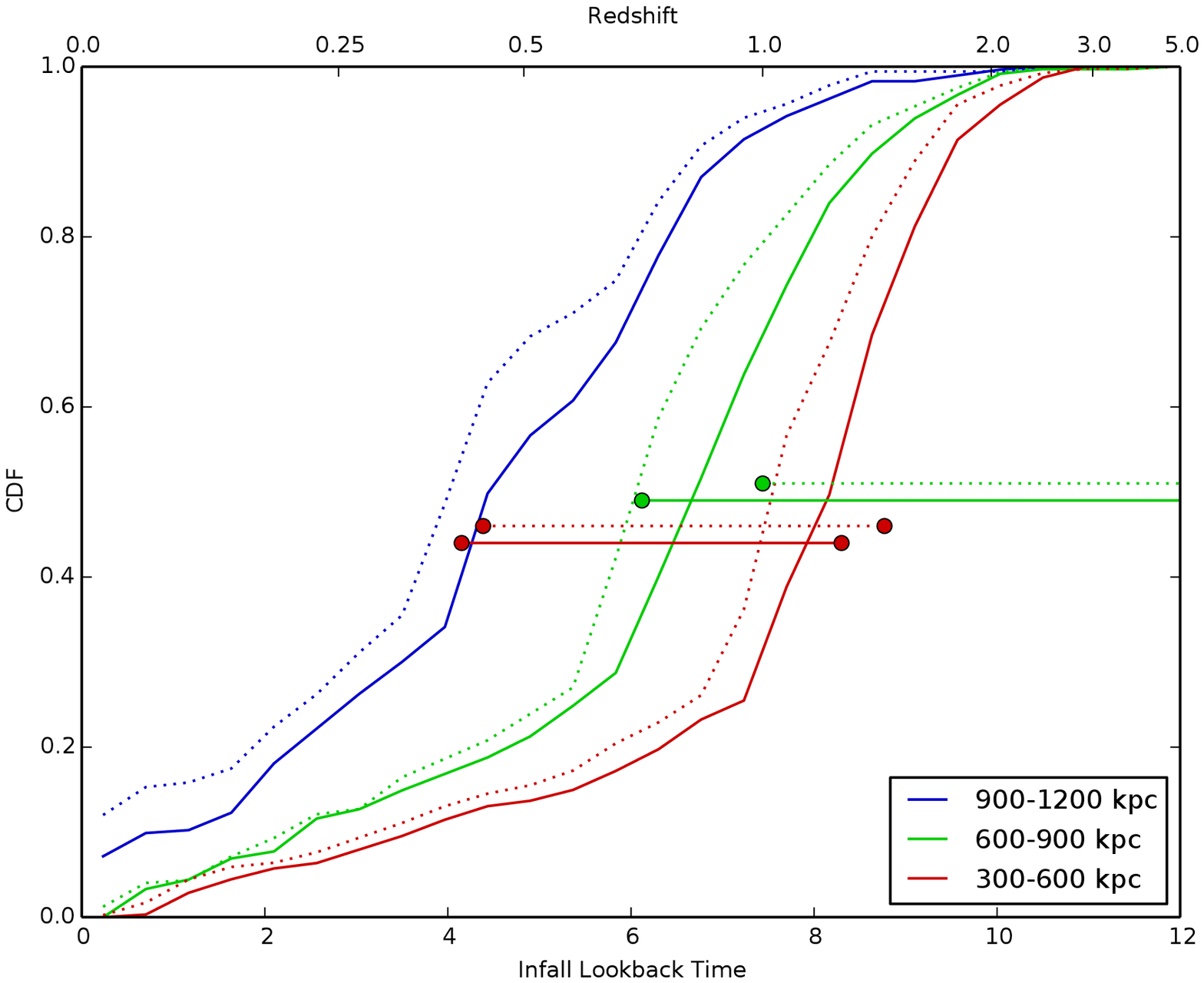}}
\caption{(Left) Stacked profile for subhalos with oldest quartile of $z_{\rm sub}$, the redshift when each object became a subhalo within any host, not necessarily its current ($z=0$) host.  The 3 different curves show stacked profiles at 3 different cluster-centric radii. (Middle) Stacked 2D surface density profiles around oldest quartile of $z_{\rm sub}$, as a function of projected radius.  (Right) For the same sample, the cumulative distribution of $z_{\rm acc}$, the redshift when each subhalo entered its current ($z=0$) host.  Solid curves correspond to 3D radial bins, and dashed curves correspond to bins of projected radius.  The horizontal bars indicate 1-2 dynamical times at each different radius.
\label{fig:sub}}
\end{figure*}

\begin{figure*}
\centerline{\includegraphics[width=0.31\textwidth]{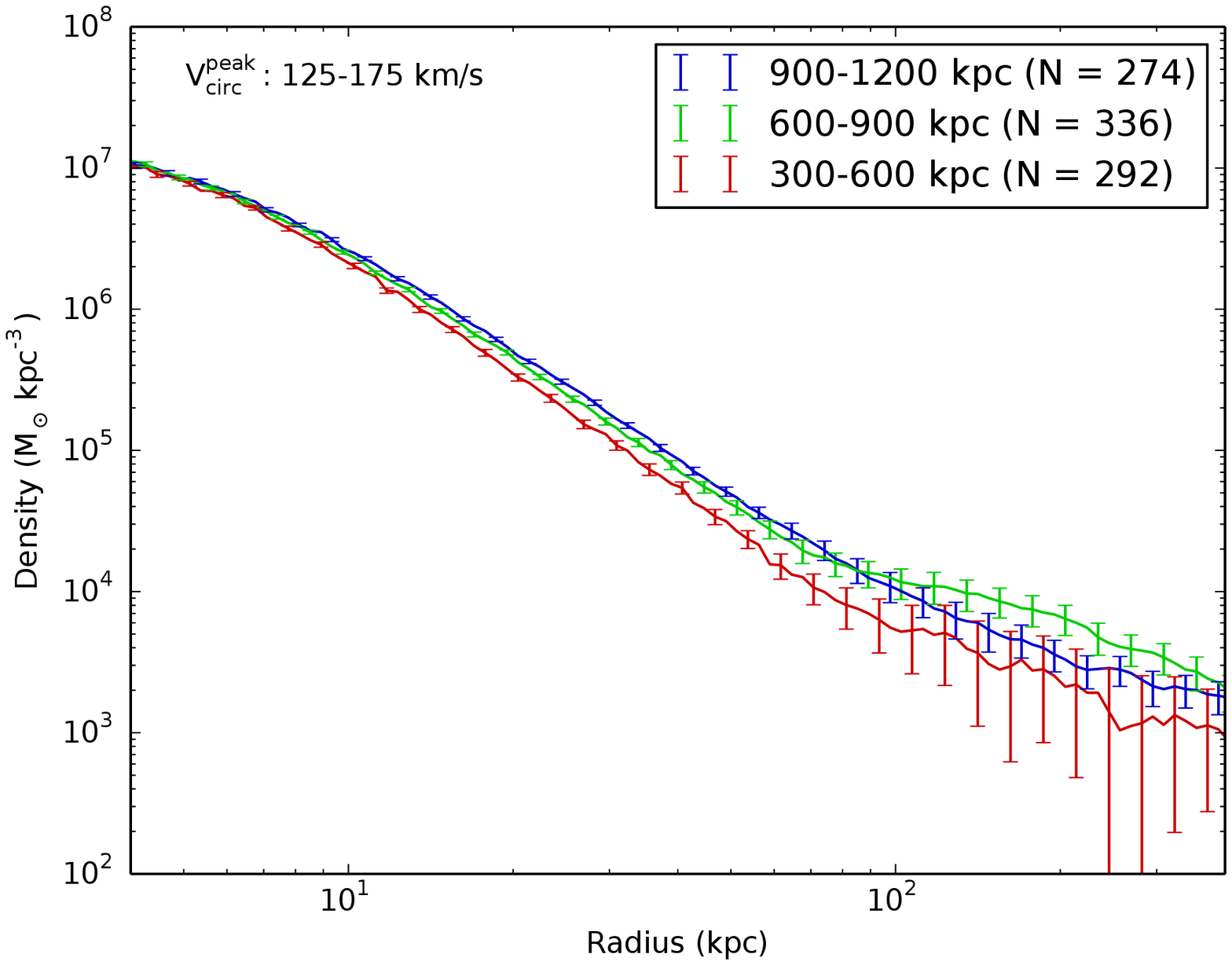}
\quad\includegraphics[width=0.31\textwidth]{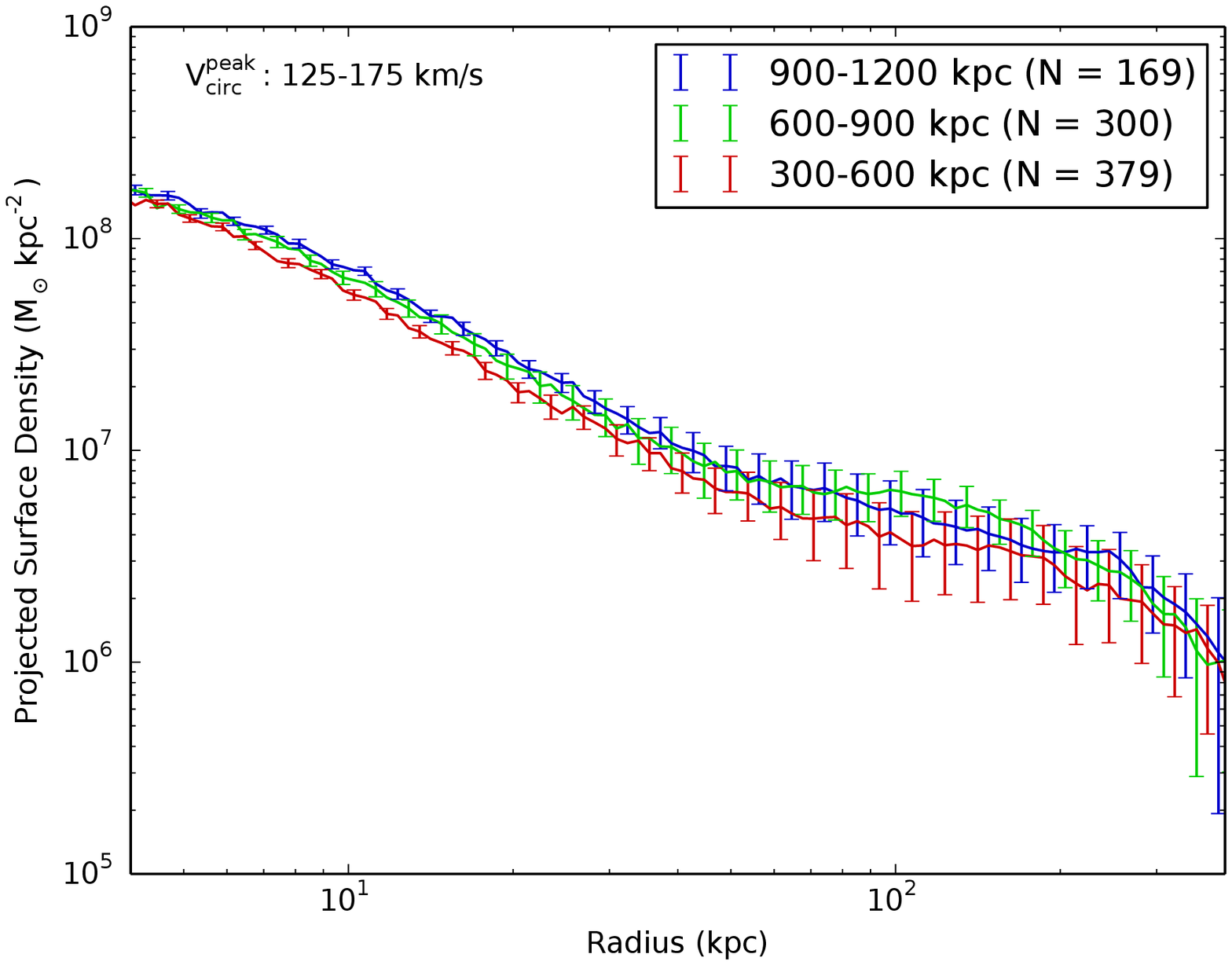}
\quad\includegraphics[width=0.31\textwidth]{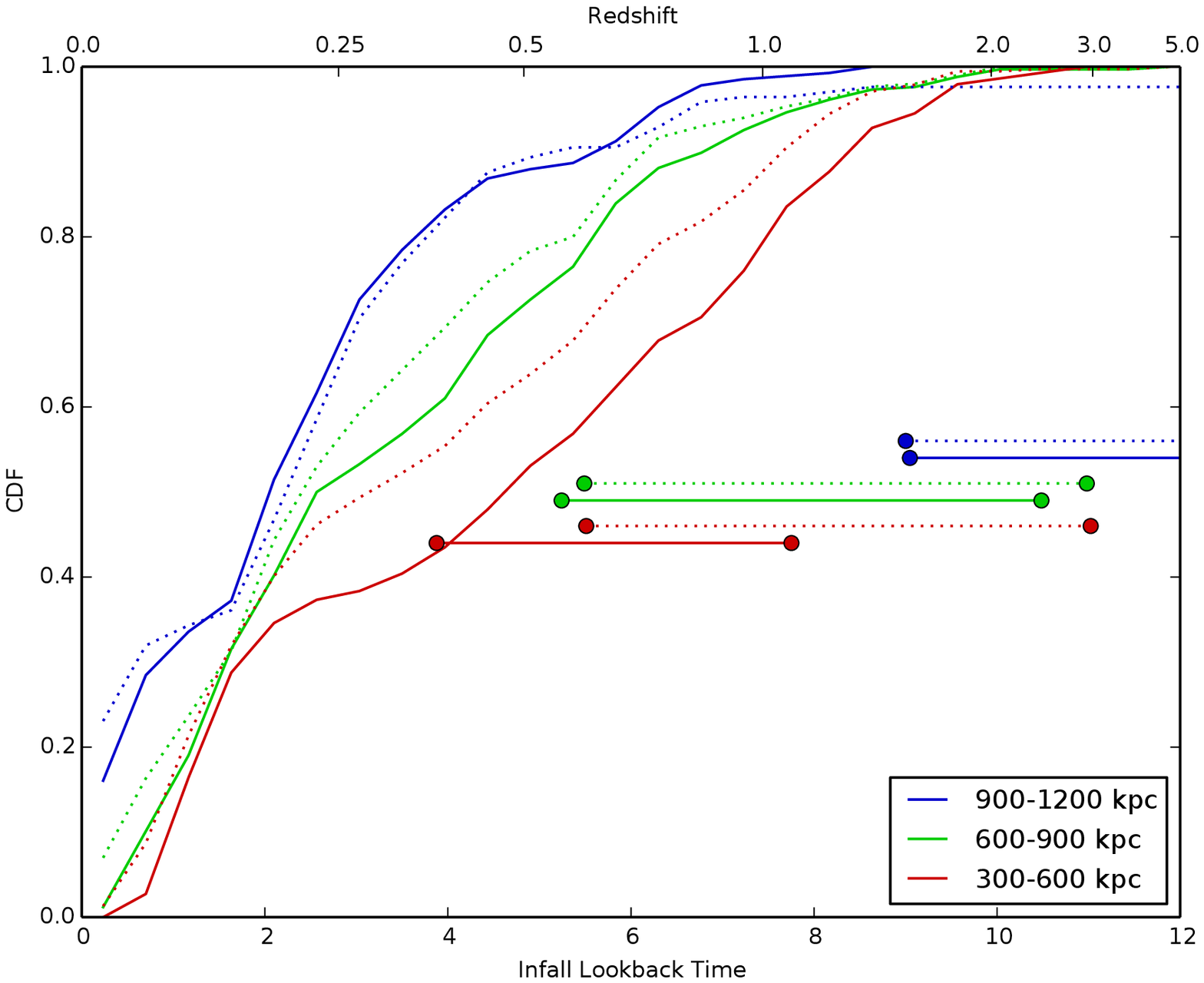}}
\caption{Same as Figure \ref{fig:sub}, but using $z_{\rm starve}$ instead.  Note that here, unlike Fig.\ \ref{fig:sub}, a large fraction of the `oldest' subhalos have been inside their current hosts for $\la t_{\rm dyn}$. Accordingly, the trends with cluster-centric radius are weaker than those in Fig.\ \ref{fig:sub}.
\label{fig:starve}}
\end{figure*}

We illustrate this point by repeating the stacking analysis of Figure \ref{fig:acc}, but for various other age estimates instead of accretion time.  We consider two examples in figures \ref{fig:sub} and \ref{fig:starve}.  In the first example (Figure \ref{fig:sub}), we use the redshift when a subhalo first enters any host ($z_{\rm sub}$) as an age estimate.  This would correspond to a model in which all quenching of star formation is associated with infall into a more massive host.  In the other example (Figure \ref{fig:starve}), we use $z_{\rm starve}$ from \citet{Hearin2013} as an age estimate.  In this model, quenching of star formation is typically not associated with infall into massive hosts.  Instead, $z_{\rm starve}$ is taken to be the oldest of either $z_{\rm W}$, $z_{\rm sub}$, or $z_{12}$, where $z_{\rm W}$ is the formation redshift defined by \citet{Wechsler2002}, determined from the subhalo's mass accretion history, and $z_{12}$ is the redshift when the object reaches a mass $M=10^{12} M_\odot$ as an isolated halo.  We determined $z_{\rm W}$ by extracting mass accretion histories from the Rockstar merger trees, and then fitting those histories to the form $M(z)\propto e^{-\alpha z}$ during the portion of the accretion history between formation of the halo and the time when it reached its peak mass.  We then translated $\alpha$ into $z_{\rm W}$ following \citet{Wechsler2002}.   For the vast majority of subhalos in our sample, $z_{\rm starve}=z_{\rm W}$.  For both age estimates ($z_{\rm sub}$ or $z_{\rm starve}$), we select the 25\% oldest subhalos at each radius, and compute their stacked profiles as in Figure \ref{fig:acc}.  The figures also show the distributions of infall redshifts, to indicate the amount of time that each sample has spent in their $z=0$ hosts.  

The figures illustrate the behavior we would expect.  At large cluster-centric radii, the subhalos have typically not resided in their present-day hosts for enough time to dissipate any correlations with unbound material.  At small cluster-centric radii ($R\sim 500$ kpc), however, the dynamical time is short enough for tidal effects to become noticeable, and the two different models produce significantly different behavior.  If quenching is associated with infall (Figure \ref{fig:sub}), then the correlations with unbound material should be nearly absent for the oldest galaxies. On the other hand, if satellite quenching is not associated with infall (Figure \ref{fig:starve}), then a large fraction of even the oldest galaxies have not resided in their current hosts for enough time to destroy correlations with unbound material.

In practice, 3D density profiles cannot be observed around galaxies.  Instead, we measure projected profiles.  The projection from 3 dimensions to 2 can wash out some of the signal, since satellites at large 3D radii can occasionally project onto small 2D radius.  Despite this mixing of different radial bins, the difference in behavior of various quenching models survives even in the projected 2D profiles (compare middle panels of Figures \ref{fig:sub} and \ref{fig:starve}).  This demonstrates that spatial clustering of satellites may be used to constrain the amount of time that a population of satellites has resided inside of their current hosts, which can help to distinguish satellite quenching models.  

In figures \ref{fig:sub} and \ref{fig:starve}, we only plot stacked mass profiles, because the small number of subhalos in the Bolshoi simulation limits our ability to measure subhalo-subhalo clustering.  Observationally, however, upcoming surveys like DES and HSC will observe much larger volumes than that simulated in Bolshoi, meaning that they will detect much larger numbers of satellite galaxies.  For example, most of the figures in this paper were generated by stacking samples of $\sim 300$ subhalos, while for comparison, the existing SDSS DR8 redMaPPer catalog \citep{Rykoff2014} contains easily two orders of magnitude as many satellites with similar host mass and cluster-centric radii as the samples used here.  Therefore, it is quite likely that the clustering signal discussed in this paper will be measured with much higher precision using real surveys than what we have measured using simulations.

\section{Discussion}

We have shown that measurements of satellite clustering can reveal the accretion histories of those satellites.  Subhalos that have orbited inside their current host halos for $\gtrsim 1-2$ dynamical times do not exhibit significant correlations with unbound material, in contrast to subhalos that entered their hosts more recently.  Since different models for satellite quenching predict different accretion histories for quenched satellites, measurements of satellite clustering can therefore constrain quenching models.

We have also argued that existing and upcoming imaging surveys will have the statistical sensitivity to measure the clustering signal proposed in this paper.  For example, the survey volumes of SDSS and DES dwarf the volume simulated in the Bolshoi simulation used in this paper.  It remains to be seen, however, if broad band photometry is sufficient to determine the ages of the stellar populations of galaxies in cluster fields, or if spectra will be required to select appropriate samples.  For example, if we are unable to distinguish quenched galaxies from dusty, reddened star-forming galaxies, then difference in the the stacked profiles of supposedly `old' and `young' galaxies will be diminished.   Line of sight interlopers present a related concern.  Because they have not experienced significant tidal gravitational forces, interlopers could dilute the stacked signal if they are mistaken for old cluster members.  Spectroscopic data could help reduce contamination by interlopers, and provide more accurate age estimates.

The observation proposed here involves measurement of 2-point correlations of satellites within clusters, and is therefore a 3-point correlation function (the cluster-satellite-satellite 3-point function).  As noted above, this is a distinct measurement from clustering statistics previously used to constrain quenching models, like average radial profiles or abundances, which involve 2-point and 1-point functions. With the advent of deep imaging surveys like DES, HSC and LSST, the measurement of 3-point and higher correlation functions is now becoming practical over a range of spatial scales.  Our estimates indicate that ongoing surveys like DES and HSC should be able to provide significant constraints on quenching models of the form examined here, which can complement other probes previously considered in the literature \citep{Cohn2014}.  In this work, we have considered only one particular sum over possible triangles, but there is clearly additional information encoded in galaxy $N$-point functions beyond the tidal effects that we have focused on. 

%{\bf TBD: extend out to larger radii to get interlopers.  maybe stack on fixed $t_{\rm dyn}$ instead of bins of $M$, $R$, etc., to improve statistics?}

\section*{Acknowledgments}
We thank Joanne Cohn, Surhud More, Eli Rykoff and Risa Wechsler for helpful discussions.  
ND is supported by NASA under grants NNX12AD02G and NNX12AC99G, and by a Sloan Fellowship.
The MultiDark Database used in this paper and the web application providing online access to it
were constructed as part of the activities of the German Astrophysical Virtual Observatory as result
of a collaboration between the Leibniz-Institute for Astrophysics Potsdam (AIP) and the Spanish
MultiDark Consolider Project CSD2009-00064. The Bolshoi and MultiDark simulations were run on the
NASA's Pleiades supercomputer at the NASA Ames Research Center.

\end{document}